\documentclass[12pt]{article}

\usepackage{graphicx}
\usepackage{amsmath}
\usepackage{amsthm}
\usepackage{amsfonts}
\usepackage[utf8]{inputenc}
\usepackage{latexsym}

\newtheorem{teor}{Theorem}[section]

\newtheorem{cor}{Corollary}[section]
\newtheorem{obs}{Remark}[section]
\newtheorem{defin}{Definition}[section]
\newtheorem{exem}{Example}[section]

\newfont{\Mb}{msbm10}
\newcommand{\C}{\mbox{\Mb\symbol{67}}}
\newcommand{\R}{\mbox{\Mb\symbol{82}}}
\newcommand{\N}{\mbox{\Mb\symbol{78}}}

\begin{document}
\setcounter{equation}{0}
\setcounter{figure}{0}
\setcounter{table}{0}

\hspace\parindent
\thispagestyle{empty}

\bigskip
\bigskip
\bigskip

\begin{center}
{\LARGE \bf A New S-Function Method}
\end{center}
\begin{center}
{\LARGE \bf searching for First Order}
\end{center}
\begin{center}
{\LARGE \bf Differential Integrals:}
\end{center}
\begin{center}
{\LARGE \bf Faster, Broader, Better.}
\end{center}

\bigskip

\begin{center}
{\large
$^a$L.G.S. Duarte, $^a$L.A.C.P. da Mota, I.S.S. Nascimento,\footnote{E-mails: lgsduarte@gmail.com and lacpdamota@gmail.com}
}

\end{center}

\bigskip
\centerline{\it $^a$ Universidade do Estado do Rio de Janeiro,}
\centerline{\it Instituto de F\'{\i}sica, Depto. de F\'{\i}sica Te\'orica,}
\centerline{\it 20559-900 Rio de Janeiro -- RJ, Brazil}

\bigskip
\bigskip
\bigskip
\bigskip

\abstract{Here we present a very efficient method to search for Liouvillian first integrals of second order rational ordinary differential equations (rational 2ODEs). This new algorithm can be seen as an improvement to the S-function method we have developed \cite{Noscpc2019}. Here, we show how to further use the knowledge of the S-function to find an integrating factor of a set of first order rational ordinary differential equations (rational 1ODEs) which is shared by the original 2ODE, without having to actually solving these 1ODEs. This new use of the S-function, that is the theoretical basis of our new method to compute the integrating factor, proved to be a linear process of computation for a vast class of non-linear rational 2ODEs, making it much more efficient.
}

\bigskip
\bigskip
\bigskip
\bigskip
\bigskip
\bigskip



\newpage

\section{Introduction}
\label{intro}

The search for methods that provide Liouvillian first integrals of polynomial
vector fields is an old problem and, paradoxically, very current in the sense that there is still much to be understood about the subject and much improvement to be made in existing algorithms.
A big step in this development was taken about forty years ago with the works of M. Singer \cite{PrSi,Sin}, which provoked a great revival in the study of methods and algorithms for obtaining Liouvillian first integrals of polynomial vector fields (see also the works \cite{Chr,Nosjpa2002-2,ChGiGiLl,Nosjcam2005,ChLlPaWa1,ChLlPe,Noscpc2007,FeGi,Che,ChLlPaWa4,LlZh,BoChClWe,Zha,ChCo,ChLlPaWa5,Dem,Nosjde2021,DeGiVa} and references therein).

We have been working on developing methods and algorithms for quite some time now. In particular, we defined a concept (the {\it S-function}, please see \cite{Nosjpa2001}) that allowed us to search for elementary first integrals of polynomial vector fields associated with rational 2ODEs, by developing an extension of the Prelle-Singer method \cite{PrSi} (see \cite{Nosjmp2009,Nosjpa2010}). This extension had a possible practical obstacle: the calculation of Darboux polynomials (DPs) in three variables was computationally `prohibitive' for degrees greater than or equal to three\footnote{The calculation of DPs is the `Achilles heel' of Darbouxian approaches, since DPs are the building blocks for integrating factors and are, generally, difficult to determine for relatively high degrees.}. To overcome this problem we created a method to search for Liouvillian first integrals of rational 2ODEs (called S-function method \cite{Noscpc2019}) that, in short, instead of determining the DPs that appear in the integrating factor, it looks for a rational function (the S-function) which defines a rational 1ODE (in fact, three) whose general solution is directly related to the first integral of the 2ODE\footnote{As the search for the S-function proved to be much more efficient in practice than the determination of the DPs, this procedure far surpassed the method developed in \cite{Nosjmp2009}.}. However, in that scenario, we still have a rational 1ODE to solve, and although this problem is much simpler than finding a first integral of a 2ODE, it can still be a considerable computational task for many interesting cases. Studying the application of our approach to linear 2ODEs we could envision a linear way of dealing with the problem that did not depend on the linearity of the original 2ODE and so it can be equally applied to the non-linear rational 2ODEs as well. Our main result in this paper is to establish a procedure that can be applied to look for Liouvillian first integrals for nonlinear rational 2ODEs linearly.

In section 2, we develop some results based on a new use of the S-function to compute Darboux polynomials in three variables linearly. In section 3 we use these results to construct integrating factors in a very efficient way. We also present a clarifying example.

\section{Computing Darboux polynomials linearly}
\label{cdpl}

In this section we will show that, curiously, a method built to avoid the (usually costly) calculation of DPs (the S-function method \cite{Noscpc2019}) turned out to be a very useful tool to accomplish this task. The S-function method, in a nutshell, consists of finding a rational 1ODE whose general solution is related to the first integral of the rational 2ODE. The method is quite effective, it trades the problem of computing DPs in three variables for that of solving a rational 1ODE. In some cases, it can be a difficult (or impossible) task either or, at least, computationally expensive. In order to better this situation, studying a possible method for dealing with linear 2ODEs, we find an approach  to use the S-function in order to determine the DPs in an alternative way. The key point was realising that, essentially,  two facts were responsible for the efficiency of the method for linear 2ODEs:

\noindent
1 - The S function (definition below) was known {\it a priory}.

\noindent
2 - There exists an integrating factor in which only one of the DPs was impossible (or hard) to find and, therefore, unknown.

None of theses facts relied on the 2ODE being linear! {\bf So we have developed a method to deal with non-linear rational 2ODEs:}

In the following subsection, we present an outline of the S-function method and, in the following one (\ref{sfmn}), applying these considerations (in other words, the knowledge of the S function and the determination of some easy to find Darboux polynomials), results that allow us to completely determine the Liouvillian First Integral for the non-linear 2ODE.

\subsection{The S-function method in a nutshell}
\label{sfmn}

\begin{defin}
\label{mathX}
Some definitions:
\begin{enumerate}
\item Consider a rational 2ODE
\begin{equation}
\label{r2ode}
z' = {\frac{M_0(x,y,z)}{N_0(x,y,z)}} = \phi(x,y,z), \,\,\,\, (z \equiv y'),
\end{equation}
where $M_0$ and $N_0$ are coprime polynomials in $\C[x,y,z]$. Let $L$ be a Liouvillian field extention\footnote{For a formal definition of Liouvillian field extention see \cite{Dav}.} of $\C(x,y,z)$. A function $I(x,y,z) \in L$ is said to be a {\bf Liouvillian first integral} (LFI) of the rational 2ODE if $\mathfrak{X}(I)=0$, where $\mathfrak{X} \equiv N_0\, \partial_x + z\,N_0\, \partial_y + M_0\, \partial_z$ is the {\bf vector field associated} (or the {\bf Darboux operator associated}) with the 2ODE.

\item Let $p\,\in\, \C[x,y,z]$ be a polynomial such that $\mathfrak{X}(p)=q\,p$. Then $p$ is said to be a {\bf Darboux polynomial} of the vector field $\mathfrak{X}$ and $q$ is a polynomial in $\C[x,y,z]$ which is called {\bf cofactor} of $p$.

\item The functions defined by $S_{k} := I_{x_i}/I_{x_j}\,$ where $i,j,k \in \{1,2,3\},\,i<j,\,k \notin \{i,j\},\,x_1=x,\,x_2=y,\,x_3=z$, are called {\bf S-functions} associated with the {\em 2ODE (\ref{r2ode})} through the LFI $I$.

\item The 1ODEs defined by ${dx_j}/{dx_i}= - S_{k}$, where $i,j,k \in \{1,2,3\},\,i<j,\,k \notin \{i,j\},\,x_1=x,\,x_2=y,\,x_3=z$ and $x_k$ is taken as a parameter, are called {\bf associated 1ODEs}  ({\em 1ODE}$_{\mathbf{[k]}} ,\, (k=1,2,3)\,$) with the {\em 2ODE (\ref{r2ode})} through $I$.
\end{enumerate}
\end{defin}

\begin{teor}
\label{teoresses}
Let $I \in L$ be a LFI of the rational {\em 2ODE (\ref{r2ode})} and let $S_{k}\,(k=1,2,3)$ be the $S$-functions associated with it through $I$.
Then

\noindent
(i) $\,I(x,y,z)=C$ is a general solution of the {\em 1ODEs} associated with the {\em 2ODE (\ref{r2ode})} through $I$.

\noindent
(ii) $\,S_{1},\,S_{2}$ and $S_3$ satisfy the following {\em 1PDEs}:
\vspace{-2mm}
\begin{eqnarray}
D_x(S_1) &=& {S_1}^2+\phi_{z}\,{S_1}-\phi_y, \label{eqs1}\\
D_x(S_2) &=& -{S_2}^2/z+\left(\phi_z-{\phi}/{z}\right)\,S_2-\phi_x, \label{eqs2}\\
D_x(S_3) &=& -{\phi_y}{S_3}^2/{\phi}+({\phi_x}-z\,\phi_y)\,S_3/{\phi}+z\,\phi_x, \label{eqs3}
\end{eqnarray}
where $D_x \equiv \partial_x + z\, \partial_y + \phi\, \partial_z\,\,\, \left( i.e.,\,D_x = \frac{\mathfrak{X}}{N_0} \right)$ is the total derivative $\frac{d}{dx}$ over the solutions of the {\em 2ODE (\ref{r2ode})}.
\end{teor}
\noindent
{\it Proof.} For a proof see \cite{Noscpc2019}.

\begin{obs}
\label{sfunm}
Remember that the S-function method is, in short\footnote{For details see \cite{Noscpc2019}.}:
\begin{enumerate}
\item $\!\!$Compute a rational solution (a rational S-function) of one of the {\em 1PDEs above (\ref{eqs1},\ref{eqs2},\ref{eqs3})}.
\item $\!\!$Solve the (corresponding) 1ODE associated to the rational {\em 2ODE (\ref{r2ode})}.
\item $\!\!$Use the general solution of the 1ODE to construct the LFI of the {\em 2ODE (\ref{r2ode})}.
\end{enumerate}
However, as it is true for the Lie method, that does not provide an algorithm for the step of finding the symmetries, the original S-function method does not prescribe any way to carry out the second step above. Let us improve that state of affairs.
\end{obs}

\subsection{Theoretical basis of our algorithm: further use of the S-function}
\label{sfmn}

\begin{defin}
\label{ourlfi}
Let $I$ be a LFI of the rational {\em 2ODE (\ref{r2ode})} and consider that its derivatives can be written as $\,I_x=R\,Q,\,\,I_y=R\,P,\,\,I_z=R\,N,$ where $R$ is a Liouvillian function of $(x,y,z)$ and $Q,\,P,\,N$ are coprime polynomials in $\C[x,y,z]$. Then we say that $I$ is a member of the set \mbox{\boldmath $L_S$} and that $R$ is an {\bf integrating factor} for the polynomial 1-form defined by $\gamma \equiv Q\,dx + P\,dy + N\,dz$.
\end{defin}

\begin{obs}
It is important to remind the reader that: $S_{k} := I_{x_i}/I_{x_j}\,$ imply that $S_1=P/N$, $S_2=Q/N$ and $S_3=Q/P$, i.e., if $I \in L_S$ then the $S_{k}$ are rational functions. As we hope to make clear just below, it is through these three rational $S_{k}$ that the determination of the integrating factor $R$ can be performed without having to solve the corresponding 1ODE (the above mentioned step 2 in remark \ref{sfunm}) and also without the problem of computing high degree DPs in three variables using the method of undetermined coefficients (MUC).
\end{obs}

\begin{teor}
\label{teoxis}
Let $I \in L_S$ be a first integral of the rational {\em 2ODE (\ref{r2ode})} such that its derivatives are written as described in the definition {\em \ref{ourlfi}}. Then, the following statements hold:

\vspace{2mm}
\noindent
$(i)\,\,$ The plane polynomial vector fields defined by $\,\mathfrak{X}_1 \equiv N \, \partial_y - P \, \partial_z, \,\,\,\mathfrak{X}_2 \equiv -N \, \partial_x + Q \, \partial_z, \,\,\,\mathfrak{X}_3 \equiv P \, \partial_x - Q \, \partial_y$, present $I$ as first integral, i.e., $\,\,\mathfrak{X}_1(I)=\mathfrak{X}_2(I)=\mathfrak{X}_3(I)=0,$

\vspace{2mm}
\noindent
$(ii) \,\,\displaystyle{\frac{\mathfrak{X}_i(R)}{R}} = - \langle \nabla , \mathfrak{X}_i \rangle  \,\,\, (i \in \{1,2,3\}).$\footnote{In what follows, the operators $\,\nabla(.),\, \langle \nabla, . \rangle,\, \nabla \wedge .\,$ stand for {\bf grad, div, culr}, respectively.}
\end{teor}

\noindent
{\it Proof.} $(i) \,\,$ The statement ({\it i}) follows directly from the definition: $\mathfrak{X}_1(I)=N \, \partial_y(I) - P \, \partial_z(I) = N\,R\,P -P\,R\,N =0$; $\mathfrak{X}_2(I)=-N \, \partial_x(I) + Q \, \partial_z(I) = -N\,R\,Q + Q\,R\,N =0$; $\mathfrak{X}_3(I)=P \, \partial_x(I) - Q \, \partial_y(I) = P\,R\,Q - Q\,R\,P =0.$

\medskip

\noindent
$(ii) \,\,$ We have that $\,\nabla \wedge \nabla(I) = 0$. This implies that

\noindent
$\partial_y(R\,N)\! - \!\partial_z(R\,P) \!=\! R_y\,N+R\,N_y - R_z\,P- R\,P_z \!= \mathfrak{X}_1(R) + R\,(N_y-P_z) = 0$,

\noindent
$\partial_z(R\,Q)\! - \!\partial_x(R\,N) \!=\! R_z\,Q+R\,Q_z - R_x\,N- R\,N_x \!=\mathfrak{X}_2(R) + R\,(Q_z-N_x) = 0$,

\noindent
$\partial_x(R\,P)\! - \!\partial_y(R\,Q) \!=\! R_x\,P+R\,P_x - R_y\,Q- R\,Q_y \!= \mathfrak{X}_3(R) + R\,(P_x-Q_y) = 0. \,\,\,\Box$

\begin{teor}
\label{rdarb}
Let $I \in L_S$ be a first integral of the rational {\em 2ODE (\ref{r2ode})}. Then the 3D polynomial vector field $\,\mathfrak{X}$ (associated with it) has a Darboux integrating factor $R$ which is also an integrating factor for the vector fields $\mathfrak{X}_i$ (for any $i \in \{1,2,3\}$).
\end{teor}

\noindent
{\it Proof.}
From the hypothesis ($I \in L_S$) and from $(ii)$ of theorem \ref{teoxis} $\left(\!\frac{\mathfrak{X}_i(R)}{R} = - \langle \nabla , \mathfrak{X}_i \rangle \!\right)$ it follows directly that the plane polynomial vector fields $\mathfrak{X}_1, \mathfrak{X}_2, \mathfrak{X}_3$ admit $R$ (an integrating factor for the vector field $\mathfrak{X}$) as an integrating factor. The Singer-Christopher results (see \cite{Sin,Chr,Nosjpa2002-2}) implies that the vector fields $\mathfrak{X}_i$ (for any $i \in \{1,2,3\}$) admit Darboux functions $R_i$ as Darboux integrating factors. So, we can write $R_i= {\cal F}_i(I)\,R\,$, where ${\cal F}_i(I)$ are functions of the first integral $I$. Therefore, the $R_i$ are also integrating factors for the vector field $\mathfrak{X}$. Since each $R_i$ is a Darboux integrating factor (in one of the pairs of variables $(x,y)$, $(x,z)$ or $(y,z)$) and all are integrating factors for the vector field $\mathfrak{X}$, then $R_i= {\cal F}_{ij}(I)\,R_j$, where ${\cal F}_{ij}(I)$ are functions of the first integral $I$. This implies that either there is a Darboux first integral since ${\cal F}_{ij}(I)=\frac{R_i}{R_j}\,$, in which case there is (certainly) a Darboux integrating factor, or ${\cal F}_{ij}(I)=k_{ij}$ (where $k_{ij}$ are constants), i.e., $R_i= {k}_{ij}\,R_j$ implying that the $R_i$ are Darboux functions on the three variables $(x,y,z)$, in fact, just one function that is a Darboux integrating factor for the vector field $\mathfrak{X}$.$\,\,\,\Box$

\begin{cor}
\label{alldps}
All DPs of $\mathfrak{X}$ present in the integrating factor $R$, are also DPs of $\mathfrak{X}_i$ ($i \in \{1,2,3\}$).
\end{cor}

\noindent
{\it Proof.}
Since $R$ is also an integrating factor for the vector fields $\mathfrak{X}_i$ ($i \in \{1,2,3\}$), the conclusion follows directly.

\begin{teor}
\label{onehardarboux}
Let $I \in L_S$ be a first integral of the rational {\em 2ODE (\ref{r2ode})} and let $\,\mathfrak{X}$ be the polynomial vector field associated with it. Also, let $R={\rm e}^{A/B}\,\prod_i {p_j}^{n_j}$, $A,\,B,\,p_j \, \in \C[x,y,z]$, $B,\,p_j$ are DPs ($p_j$ irreducible), $n_j$ are rational numbers, be a Darboux integrating factor which is also an integrating factor for the vector fields $\mathfrak{X}_i$ ($i \in \{1,2,3\}$). If only one irreducible DP, say $p_0$, is unknown and it is not part of $B$, then $p_0$ can be determined by solving linear algebraic systems.
\end{teor}

\noindent
{\it Proof.}
From the hypotheses of the theorem
\begin{equation}
\label{xirsr}
\frac{\mathfrak{X}_i(R)}{R} = \mathfrak{X}_i\!\left(\frac{A}{B}\right)+\sum_j\,n_j\,\underbrace{\frac{\mathfrak{X}_i(p_j)}{p_j}}_{q_{ij}} = - \langle \nabla , \mathfrak{X}_i \rangle.
\end{equation}
Since $\mathfrak{X}_i(A/B)\!=\!(B\,\mathfrak{X}_i(A)-A\,\mathfrak{X}_i(B))/{B^2}$ is polynomial and $B$ can be determoned linearly (see \cite{Nosjcam2005,Noscpc2007}), the equations $B\,\mathfrak{X}_i(A)$ $-A\,\mathfrak{X}_i(B)=B^2\,{\cal P}$ are linear in the unknown coefficients. So, we can find $A$ and ${\cal P}$ linearly. Denoting $p_0$ as the unknown DP, we can write the equations (\ref{xirsr}) as ${\cal P} + \sum_j n_j\,q_{ij} + n_0\,q_{i0} +\langle \nabla , \mathfrak{X}_i \rangle=0$ which are linear in the coefficients of the polynomial $n_0\,q_{i0}$ and in the exponents $n_j$. Once determined, we can use the equations $\mathfrak{X}_i(p_0)=q_{i0}\,p_0$ (linear in the coefficients of $p_0$) to determine $p_0$.$\,\,\,\Box$


\section{An efficient method to search for LFI}
\label{emslfi}

Regarding the results shown in the last subsection, we can highlight two important points:
\begin{obs} \
\label{impres}
\begin{itemize}
\item The DPs $p_j(x,y,z)$ of $\mathfrak{X}$ that compose the integrating factor $R$ are, with relation to vector fields $\mathfrak{X}_i$, DPs in two variables  (and, in general, with a lesser degree). Ex. the DP $p=x^3\,z-y$ in three variables, has degree 4 with respect to the vector field $\mathfrak{X}$, however, with respect to $\mathfrak{X}_1$ ($\equiv N \, \partial_y - P \, \partial_z$), it is a DP in two variables with degree 1.
\item It makes no difference to the method whether the unknown DP is irreducible or not. In this way, any unknown polynomial in the integrating factor {\bf can be found linearly} with the same strategy.
\end{itemize}
\end{obs}

\subsection{The New algorithm}
\label{aes}


\noindent
Remark \ref{impres} above give us a clue to establish an efficient strategy to determine the DPs that are factors of an integrating factor of the rational 2ODE: First, we can use the vector fields $\mathfrak{X}_i$ to calculate all `easy to find' DPs (in practice, DPs of degrees 1 and 2)\footnote{These DPs can have high degree in the three variables $(x,y,z)$.}; then, we can use the equations $\mathfrak{X}_i(R)/R=- \langle \nabla , \mathfrak{X}_i \rangle$ to determine $q_0$ (assuming that there would still be an unknown DP of very high degree); finally, we could use the equations $\mathfrak{X}_i(p_0)= p_0\,q_0$ to determine $p_0$ (linearly, since $q_0$ has already been determined).
This enables us to find a procedure to perform the still unresolved part of the method (remark \ref{sfunm}, step 2), i.e., solve the associated 1ODE.

\noindent
{\bf Procedure {\it DPL} (sketch):}
\begin{enumerate}
\item Determine the DPs of degree 1 (and 2 if possible) of the vector fields $\mathfrak{X}_i$ ($i \in \{1,2,3\}$). With them (if some are found), determine possibles $A$ and $B$ and therefore, possibles ${\cal P}_i$, by solving the equations $B\,\mathfrak{X}_i(A)$ $-A\,\mathfrak{X}_i(B)=B^2\,{\cal P}_i$.

\item Construct candidates for $n_0\,q_{0i}$ and substitute then in the equations ${\cal P}_i + \sum_{j\neq 0} n_j\,q_{ij} + n_0\,q_{i0} +\langle \nabla , \mathfrak{X}_i \rangle=0$ (see proof of theorem \ref{onehardarboux}). Collect the equations in the variables $(x,y,z)$. Solve the set of equations for the $n_j$, and for the coefficients of the $n_0\,q_{0i}$ candidates.

\item Construct a candidate for $p_0$ with undetermined coefficients and substitute it in the equations $\mathfrak{X}_i(p_0)=q_{i0}\,p_0$. Collect the equations in the variables $(x,y,z)$ and solve the set of equations for the coefficients of the $p_{0}$ candidate.

\item Construct the integrating factor $R={\rm e}^{A/B}\,\prod_i {p_j}^{n_j}$ and determine the LFI $I$.

\end{enumerate}

\begin{obs}
There is a huge variety of ways to build procedures based on the points highlighted by the remark \ref{impres}. For example, in the case of a rational 2ODE that presents an elementary first integral, it is not necessary to use step 2 (${\cal P}=0$). Therefore, some variant of the method could test this condition in the first place. However, the most important point to emphasize is that the knowledge of the vector fields $\mathfrak{X}_i$ allows us to obtain DPs in a much less expensive way and, in the case that some DP of very high degree is still undetermined, it can be found by a linear process (even if it is not irreducible). In fact, except for the first part of step 1 (i.e., determining the `easy to find' DPs), all processes in steps 1, 2 and 3 are linear.
\end{obs}

\begin{exem} {Illustrative example} \label{ilustr1} \end{exem}
Consider the non-linear rational 2ODE
\begin{equation}
\label{ex1-2ode}
z'=-{\frac { \left( {z}^{3}{x}^{6}-2\,y{z}^{3}{x}^{4}-2\,yz{x}^{4}+{x}^{2
}{y}^{2}+2\,{y}^{3} \right)  \left( zx-2\,y \right) }{{x}^{5} \left( 3
\,{z}^{5}{x}^{4}+2\,{z}^{3}{x}^{4}-3\,{z}^{2}y{x}^{2}-3\,{y}^{2}{z}^{2
}+{y}^{2} \right) }}.
\end{equation}
We can determine $S_3=-2\,y/x$ and, from it\footnote{These calculations are the first step of our S-function Method, mentioned above, which is the first step of our new method as well.},
\begin{eqnarray}
\label{ex1-sfs12}
S_1&=&{\frac {{x}^{6}{z}^{3}-2\,{x}^{4}y{z}^{3}-2\,{x}^{4}yz+{x}^{2}{y}^{2}+
2\,{y}^{3}}{{x}^{4} \left( 3\,{x}^{4}{z}^{5}+2\,{z}^{3}{x}^{4}-3\,{x}^
{2}y{z}^{2}-3\,{y}^{2}{z}^{2}+{y}^{2} \right) }}, \\
S_2&=&-2\,{\frac {y \left( {x}^{6}{z}^{3}-2\,{x}^{4}y{z}^{3}-2\,{x}^{4}yz+{x
}^{2}{y}^{2}+2\,{y}^{3} \right) }{{x}^{5} \left( 3\,{x}^{4}{z}^{5}+2\,
{x}^{4}{z}^{3}-3\,{x}^{2}y{z}^{2}-3\,{y}^{2}{z}^{2}+{y}^{2} \right) }}.
\end{eqnarray}
From this point we can apply the procedure {\it DPL}\footnote{That constitutes the novelty of our new Method, that allows for the fact that or computation of the LFI can be so efficient}: in short, we can find a DP of degree 1: $p_1=x$. We find ${\cal P}=0$ (i.e., there is an integrating factor without the exponential part, which means that there can be an elementary first integral), $n_1=-1$. Following the procedure we find $n_0\,q_{03}= -8\,{x}^{6}{z}^{3}+16\,{x}^{4}y{z}^{3}+16\,{x}^{4}yz-8\,{x}^{2}{y}^{2}-16\,{y}^{3}$, leading to $p_0={x}^{4}{z}^{3 }-y^2$, $n_0=-2$ and the integrating factor $R=1/(x \left( {x}^{4}{z}^{3}-y^2 \right) ^{2})$. Finally, we get $\,I = -{{\rm e}^{{\frac { \left( {x}^{2}z-y \right) {x}^{2}}{{x}^{4}{z}^{3}-{y}^{2}}}}}{x}^{4} \left( {x}^{4}{z}^{3}-{y}^{2} \right) ^{-1}$.

\begin{obs} Some comments and considerations:
\begin{enumerate}
\item The {\em 2ODE (\ref{ex1-2ode})}, our demonstrating example \ref{ilustr1} above, is not `easy', i.e., its first integral is not easily found by any `canonical' method: it does not belong to a set of 2ODEs with a given type and known solution; it has no point symmetries; one of the DPs present in the integrating factor has degree 7. However, the procedure {\it DPL} finds the DP $p_0={x}^{4}{z}^{3 }-y^2$ almost instantly, thus demonstrating its great efficiency.
\item It is important to note that we could, in the case of {\em 2ODE (\ref{ex1-2ode})}, avoid the step of calculating the DPs (of degree 1 and 2) of the vector fields $\mathfrak{X}_i$ ($i \in \{1,2,3\}$). It would suffice to try to find directly the DP $x \left( {x}^{2}{z}^{4}-y^2 \right) ^{2}$ that appears in the denominator of the integrating factor. It could even be a product of several high degree irreducible DPs (see example \ref{ilustr2} bellow).
\item What makes the case that we can treat linearly (i.e., only one unknown DP) so comprehensive is directly related to the fact described in observation 2 (just above). That is, we can treat the entire product of polynomials that appear in the denominator (or in the numerator) of the integrating factor as just a DP of very high degree. This allows us to deal with cases where the product of DPs appearing in the numerator (resp. denominator) are of low degree in two of the three variables and the product of DPs appearing in the denominator (resp. numerator) may be the most general, in a linear or quasi-linear fashion. (see example \ref{ilustr2} bellow).
\item {\bf Summarising:} Our results presented here can be seen as the new S-method (originally presented on \cite{Noscpc2019}). That is why we presented a summary of the method on section \ref{sfmn} since its beginning is also the beginning of our method here. The point that was very effective in improving the original method \cite{Noscpc2019} was the exchanging of the solving of the associated 1ODEs for the clever usage of the S-function to determine the `low degree' DPs and that its resulting algorithm for computing the `high degree' DPs is linear (if all are in numerator or in denominator), very fast compare to other possible methods to do the task, making it, in many cases, practically feasible.
\end{enumerate}
\end{obs}

\begin{exem} {Illustrative example 2} \label{ilustr2} \end{exem}Consider the non-linear rational 2ODE
\begin{equation}
\label{ex1-2ode}
z'={\frac { \left( {z}^{12}{x}^{14}+{z}^{9}{x}^{14}+{z}^{3}{x}^{14}-
3\,y{z}^{6}{x}^{7}-y{z}^{3}{x}^{7}-{x}^{7}y+2\,{y}^{2} \right)
 \left( zx-7\,y \right) }{ 3{z}^{2}{x}^{8}\left( 4{z}^{15}{x}^{14}\!+\!{z}^{12}{x}^{14}
\!+\!2{z}^{9}{x}^{14}\!+\!{z}^{6}{x}^{14}\!-\!10y{z}^{9}{x}^{7}\!-\!4y{z}^{3}{x}
^{7}\!+\!{x}^{7}y\!+\!6{y}^{2}{z}^{3}\!-\!{y}^{2} \right) }}.
\end{equation}
For this 2ODE, $S_3=-7\,y/x$ and from it we have the vector fields $\mathfrak{X}_i$. The point is that we can find the DP $p_0=x\left( {x}^{7}{z}^{6}-y \right) ^{2} \left( {x}^{7}{z}^{6}+{x}^{7}-y \right)$ in less than $0.2 sec$ using the equation $q_0- \langle \nabla , \mathfrak{X}_2 \rangle=0$ followed by the equation $\mathfrak{X}_2(p_0) -q_0\,p_0=0$. This was possible because there exists a Darboux integrating factor $R=1/(p_0)$, i.e., the two DPs of high degree $\,{x}^{7}{z}^{6}-y \,$ and $ \,{x}^{7}{z}^{6}+{x}^{7}-y\,$ were both in the denominator of $R$. From $R$ we can determine $I={{\rm e}^{{\frac {{z}^{3}{x}^{7}-y}{{x}^{7}{z}^{6}-y}}}}{x}^{14}
 \left( {x}^{7}{z}^{6}-y \right) ^{-1} \left( {x}^{7}{z}^{6}+{x}^{7}-y
 \right) ^{-1}
$.


\end{document}